\documentclass[conference]{IEEEtran}
\usepackage{multirow}
\usepackage{graphicx}
\usepackage{caption}
\usepackage{subcaption}
\usepackage[table]{xcolor}
\usepackage{array}
\usepackage{url}
\usepackage{amsmath}
\usepackage{Definitions}
\usepackage{authblk}
\usepackage[pdfborder={0 0 0}]{hyperref}

\graphicspath{{.}}

\title{A Continuous-time Mutually-Exciting Point Process Framework for Prioritizing Events in Social Media}

\begin{document}

\author[1]{Mehrdad Farajtabar}
\author[2]{Safoora Yousefi}
\author[3]{Long Q. Tran }
\author[1]{Le Song}
\author[1]{Hongyuan Zha}

\affil[1]{Georgia Institute of Technology \\
mehrdad@gatech.edu, lsong@cc.gatech.edu, zha@cc.gatech.edu}
\affil[2]{Emory University \\
safoora.yousefi@emory.edu}
\affil[3]{VNU-Hanoi \\
tqlong@vnu.edu.vn}



\maketitle

\begin{abstract}
The overwhelming amount and rate of information update in online social media is making it increasingly difficult for users to allocate their attention to their topics of interest, thus there is a strong need for prioritizing news feeds. The attractiveness of a post to a user depends on many complex contextual and temporal features of the post. For instance, the contents of the post, the responsiveness of a third user, and the age of the post may all have impact. So far, these static and dynamic features has not been incorporated in a unified framework to tackle the post prioritization problem.

In this paper, we propose a novel approach for prioritizing posts based on a feature modulated multi-dimensional point process. Our model is able to simultaneously capture textual and sentiment features, and temporal features such as self-excitation, mutual-excitation and bursty nature of social interaction. As an evaluation, we also curated a real-world conversational benchmark dataset crawled from Facebook. In our experiments, we demonstrate that our algorithm is able to achieve the-state-of-the-art performance in terms of analyzing, predicting, and prioritizing events. In terms of interpretability of our method, we observe that features indicating individual user profile and linguistic characteristics of the events work best for prediction and prioritization of new events.

\end{abstract}

\section{Introduction}
Online social media (OSM) and communities turn to become an inseparable part of today's lifestyle. Users of OSM usually participate via a variety of ways, including but not limited to sharing text and photos, asking questions, publishing their status, finding friends, and favoring or  disfavoring contents. Many of these are organized around discussion threads which are evolving continuously. A  mechanism of publishing \emph{posts} and \emph{commenting} on posts is an essential part of many social networking websites, forums, and groups.
Users tend to pay attention to posts and comments that are from their preferred connections, or that are of topical attractiveness. However, the rate at which such discussion threads (cascades) are generated and developed is extremely high, and as a consequence, the user has to spend considerable time to find events of interest, or they may miss many appealing discussions, especially when they have many connections in the network. On the other hand, the cluttered news feed of a social network user or the cramped homepage of an online community member makes them reluctant to continue using the service. 

The \emph{news feed prioritization} problem deals with sorting the events in the a user's news-feed in a way that her stories of interest end up in the top of her newsfeed.
There is a pressing need for algorithms and tools to prioritize the news feeds of online social media and discussion groups. Despite the significance of it from the customer satisfaction point of view, few solutions have been proposed for this problem. A detailed discussion of the previous work is given in Section 4.

The concept of an ideally sorted newsfeed is usually subjective, and varies greatly from user to user. That's why a data-driven solution emerges for this problem.
Basically, the attractiveness of a discussion thread (cascade) to a user depends on many complex social, contextual and temporal features of the cascade, such as its age, its content, and the reactions of other users to it.  So far, these static and dynamic features have not been incorporated in a unified framework to tackle the news feed prioritization problem. 


In this paper, we will address this problem using a novel framework based on a feature modulated multi-dimensional point process which can simultaneously capture both static and dynamic features of social interactions. Recently, there has been a growing interest in the application of point processes to social media analysis \cite{FarWanGomLiZhaSon15,FarGomDuZamZhaSon15, zhou2013learning,yang2013mixture, farajtabar2014shaping, nankdd15}. \emph{Hawkes Process} \cite{hawkes1971spectra} is a special form of point process which exhibits \emph{mutual excitation}  and is appropriate for modeling user interactions in social media.
In this paper we use network data, and infer social influence among users of an OSM from their generated content and interaction history. However, to learn the quadratic number of parameters of the Hawkes process accurately, a large amount of interaction data in terms of both the number of sample cascades and cascade length is needed. To overcome this problem, we propose to model social influence as a combination of features extracted from the users' behavior and published content. Furthermore, our model is also easy to interpret: we can tell to what extent each feature contributes to social influence based on the parameters of the learned model.

\begin{figure*}
\centering
\advance\leftskip-1cm
\includegraphics[width=0.74\textwidth]{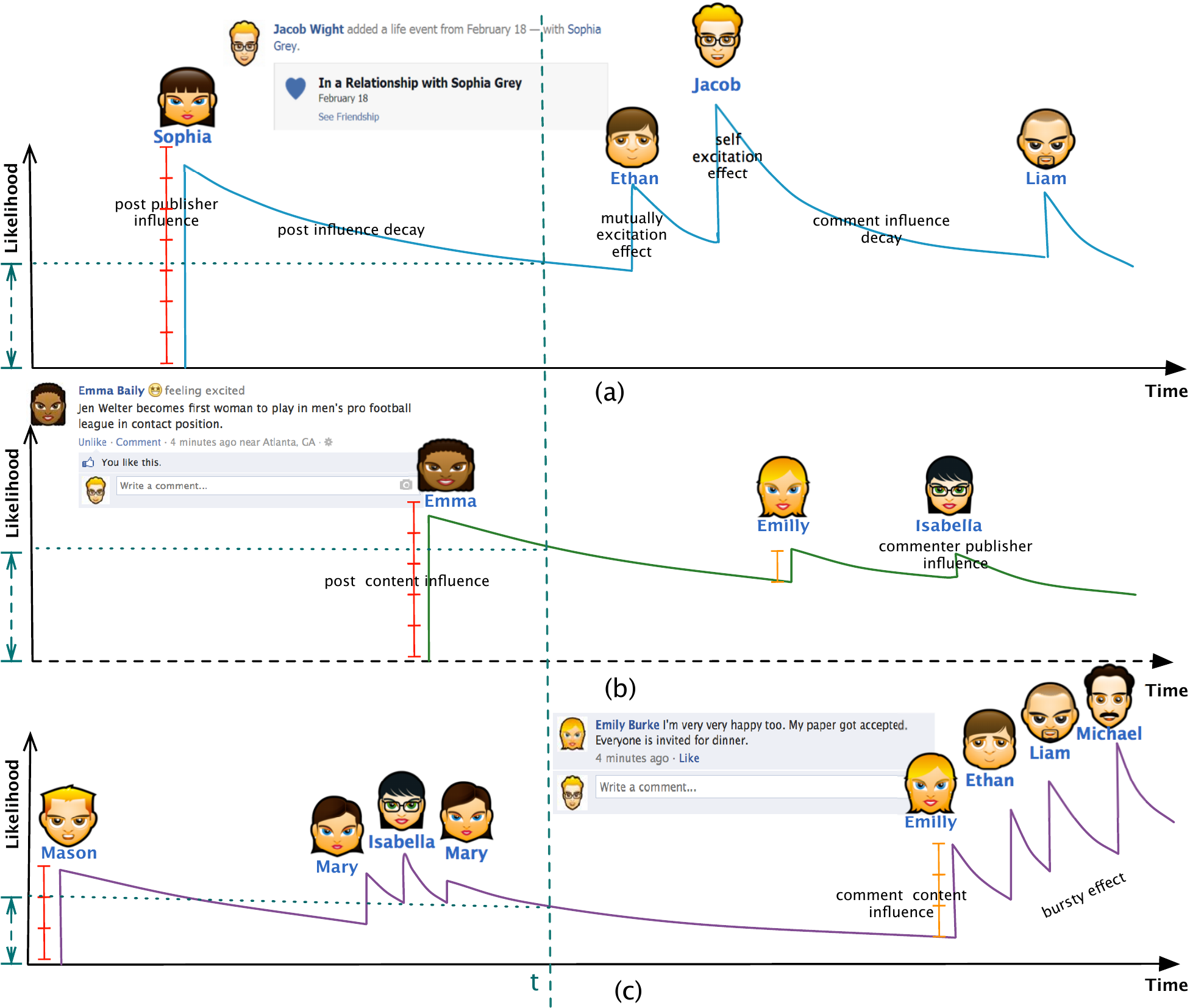}
\vspace{-2mm}
\caption{The evolution of the likelihood of Jacob's contribution to three posts by a) Sophia, b) Emma, and c) Mason with regard to time and other related events.}
\vspace{-8mm}
\label{motiv_fig}
\end{figure*}

Last but not least, in order to test the efficacy of the proposed method we curated a first-of-the-kind benchmark dataset of multi-user conversations in Facebook. The dataset contains about 50,000 posts and about 1,000,000 comments made by almost 25,000 users naturally partitioned into 16 active Facebook groups, making it an ideal dataset for evaluating news feed prioritization algorithms. To summarize, the contributions of this paper are three-fold:




\begin{itemize}[noitemsep]
\item Inspired by real-world dynamics of conversational activities, including their self- and mutual-excitation properties as well as temporal properties such as burstiness \cite{barabasi2005origin}, we propose a novel framework based on multi-dimensional Hawkes process for conversation modeling and prioritization.
\item As a solution to data scarcity, we parametrize social influence as a weighted sum of features, enabling easy interpretation of the model and the study of the contribution of each feature to social influence.
\item We introduce a new compelling conversation dataset to evaluate the proposed approach, which has recently been crawled from Facebook and is publicly available as a benchmark dataset.
\end{itemize}

The rest of the paper will first explain the intuitive dynamics of conversations in Section 2. Section 3 is the key technical section of the paper where, after building sufficient background, we present our proposed prioritization algorithm and the features designed to derive social influence. We surveyed related work in section 4. Section 5 presents the evaluation and analysis of the proposed algorithm using a real-world dataset. At last, the paper is concluded in section 6.

\vspace{-1mm}
\section{A Motivating Example}
\vspace{-1mm}
\label{motivation}

Figure \ref{motiv_fig} illustrates a detailed example by which we explain our observations on social interactions. Consider Jacob's newsfeed on the OSM. His three friends, Sophia, Emma, and Mason have recently posted status updates. As a result, three threads of conversation start. Jacob may comment on these threads with different likelihoods. These likelihoods evolve with time due to a quality called the \emph{time subordination} of events.

The initial likelihood may depend on the social influence of each of the three friends on Jacob. Sophia, for instance, is Jacob's best friend, and  they have recently become involved in a relationship. Therefore, the likelihood of Jacob commenting on her post is normally high, as demonstrated in part \textbf{a} of figure \ref{motiv_fig}. We call this property the \emph{post publisher influence}. 
On the other hand, Jacob is a fan of Football, therefore, Emma's post is on a popular topic -Football- so it excites Jacob and increases his likelihood of commenting on her post to a value higher than it would be with regard to his relationship with Emma only. As you have noticed, the content of the post effectively induces Jacob to respond to it. We refer to this effect as \emph{post content influence}. These two sources of influence are combined to form the overall \emph{post influence}.

\emph{Post influence decay} is another important factor determining likelihood. As a post ages, it is drawn out of focus by the generation of new events.  Jacob's feed is accumulated with recent scenarios that distract him from Sofia's post. 

Let us now track Sophia's post. The likelihood that Jacob pays attention to Sofia's post keeps decreasing until Ethan comments on it and brings it to the top of Jacob's news feed as a recent event. This is basically what is known as the \emph{mutual-excitation} of events in OSM. Now, let's say Jacob finally comments on Sophia's post. Interestingly, after he comments on it, his likelihood of returning to Sophia's post and re-contributing increases. This property is a special case of the previous one, and we will refer to it as the \emph{self-excitation} property.

Now, let's consider Mason's post. It seems that neither Mason has much influence on Jacob nor does his post's content attract Jacob. Therefore, the likelihood starts with a relatively low value. The post remains idle for a while and the likelihood attenuates, until Mary and Isabella comment on it, and, loosely speaking, awaken the post. Once more, the post becomes idle for a relatively long time until when Emily comments on it. Surprisingly, the post becomes very attractive afterwards, and three other people comment on it. Mutual-excitation accumulates likelihood and makes this post a hot one. This scenario reflects the \emph{bursty} nature of social events\cite{barabasi2005origin}.

There still remains an unexplained point. Before Emily comments on Mason's post, she comments on Emma's post. However, it does not attract Jacob's attention very much. Therefore, some other factor besides influence is playing a role: the content of the comment.
It can be seen that the \emph{comment content influence} is another factor driving behavior in social networks. 
On the other hand, Isabella usually has little influence on Jacob since Jacob does not like her very much, irrespective of what she expresses in her comment. This final example indicates that \emph{comment publisher influence} is another component of the \emph{comment influence}.

Similar to posts, comments normally exhibit a \emph{comment influence decay} due to the same reasons. As time passes, more attractive stories replace them. However, as it is apparent, the influence of posts will last longer. In other words, the influence decay happens faster for comments than for posts. \emph{Different rates of decay} is due to the fact that  posts are generally more observable and easier to follow.

We end our observation at some point in time. Jacob may or may not comment on Emma's and/or Mason's post at a later time.
To put it in a nutshell, conversations exhibit the following properties which we want to capture in a unified model:
\begin{figure*}
\centering
\includegraphics[width=0.9\textwidth]{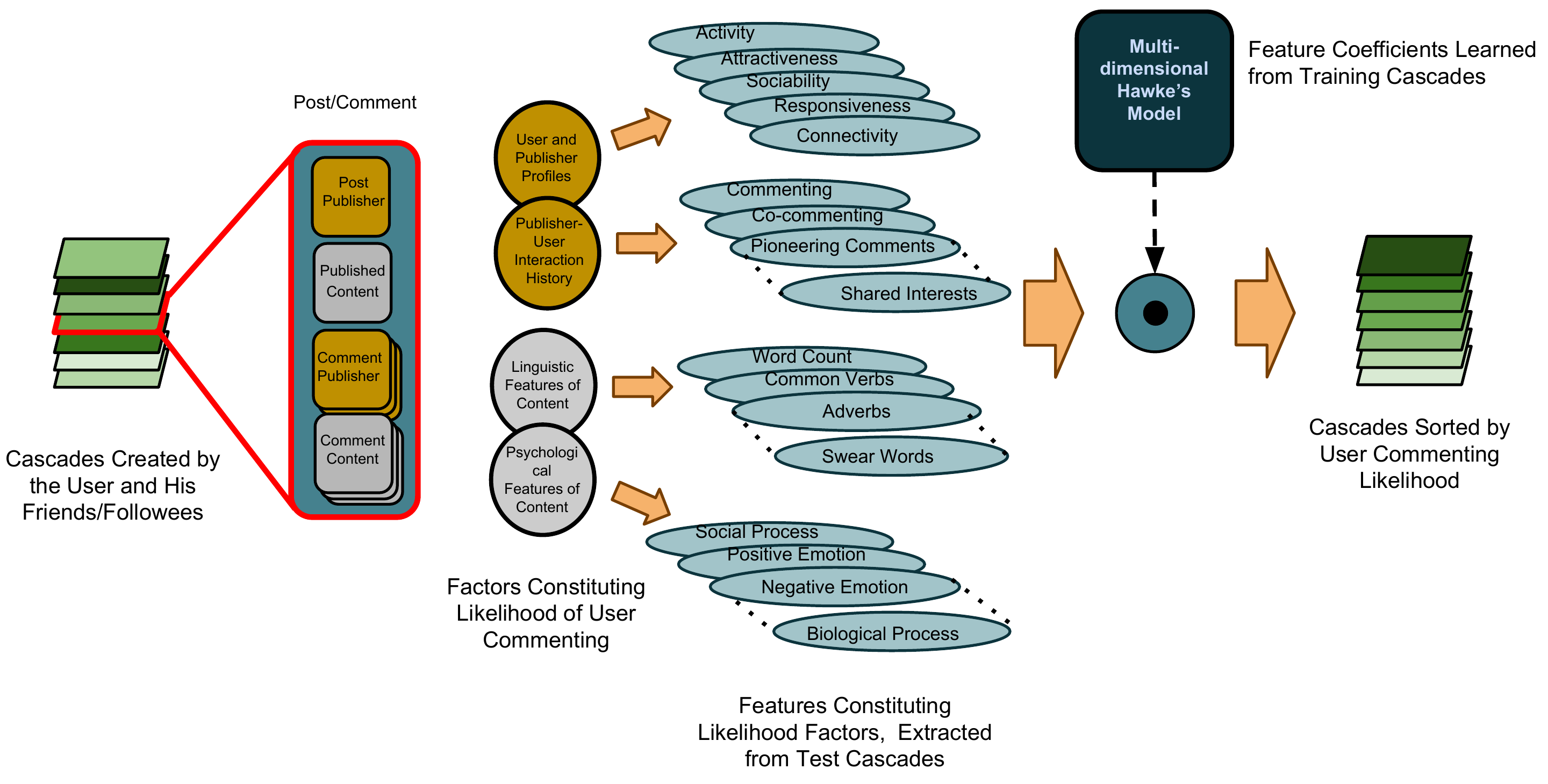}
\vspace{-2mm}
\caption{Graphical presentation of our newsfeed prioritization method.}
\label{proposed-fig}
\vspace{-6mm}
\end{figure*}

 \textbf{Time Subordination:} The likelihood of commenting evolves with time.

 \textbf{Post Publisher Influence:}  The likelihood of commenting is proportional to the social influence of the publisher on the subject user.

 \textbf{Post Content Influence:}  The likelihood of commenting is related to content of the post.

 \textbf{Post Influence Decay:} As time passes, the probability of commenting on the post decreases.

 \textbf {Comment Publisher Influence:} It reflects the intrinsic influence of the commenter on the subject user

\textbf {Comment Content Influence:} The content of the comment is an important factor triggering other events.

 \textbf {Comment Influence Decay:} As time goes on, the effect of the comment disappears gradually.

 \textbf{Different Rate of Decay:} Comment and post influences decay at different rates.
 
 \textbf {Mutual-excitation:} When a third user comments on the post, the likelihood of the subject user commenting increases.

 \textbf {Self-excitation:} When one comments on the post there is a relatively higher chance that he returns to the post and comment again.

 \textbf{Bursty Effect:} The comments on a post exhibit a bursty nature;  after a post is made, it may receive multiple comments, then become dormant  for a relatively lone time after which it comes to life again by receiving new comments.

 Returning to our problem of interest, prioritization, we can sort Jacob's news feed based on his \emph{likelihood} of commenting on posts. In our synthetic example, suppose we are at time $t$ when Jacob refreshes his page (we have not seen the events after time $t$ yet). Based on our notion of influence, Jacob is most likely to comment on Emma's post. The next attractive post to him is Sophia's post, and Mason's comes last. Therefore, we had better  show the stories to Jacob in the mentioned order. As time goes on,  new events are generated, and the likelihoods change. For example, later on, we see that after Ethan's comment on Sohpia's post, the likelihood of Jacob commenting on her post increases, and if we get Jacob's feed sorted according to the likelihood, he will hit his topic of interest at the top of his news feed.
 In the subsequent sections we elaborate on a model to capture these properties and then prioritize events based on the intensity.

\vspace{-1mm}
\section{Backgrounds on Temporal Point Processes}
\vspace{-1mm}

Our framework elaborates upon the theory of temporal point processes and event history analysis. Time is an important factor in the study of events in social networks. One might argue that any well-developed statistical method such as linear or ridge regression, probabilistic models, etc. can be used with waiting times of events to conduct an analysis of event dynamics on a sample of users. The reason these standard statistical methods do not suit time-dependent events is a fundamental problem one almost always meets when dealing with such problems: when the study of sample of users ends and the analysis begins, one is left with a set of both complete and incomplete observations. The event in question has happened for a subset of users and not for the rest, but one does not have the knowledge of whether in future the event is going to happen to the rest. \cite{aalen2008survival}. 

Temporal point processes model point patterns with stochastic processes and are suitable for dealing with incomplete observations \cite{aalen2008survival,daley2002introduction}.
The term point refers to the conceptualization of an event as instantaneous and representable as a point on the time line.
A (1-dimensional) point process is simply a list of points in time $\{t_1, t_2, ..., t_N\}$ at which events  occur. Equivalently,  a point process can be defined by the corresponding counting process denoted by $N(t)$-the number of occurrences up to time $t$.  
Conditional intensity function is the most convenient and intuitive way to characterize a point process. Defining the history of events up to but not including $t$ as $\mathcal{H}_t$, the intensity function is
\begin{equation}
\lambda(t|\mathcal{H}_t) = \lim_{\Delta t \to 0}
\frac{\mathbb{E}[N(t+\Delta t)-N(t)|\mathcal{H}_t]}{\Delta t}
\end{equation}
An intuitive interpretation of conditional intensity function is: $\lambda(t|\mathcal{H}_t) \, \Delta t$ is the
conditional probability of observing an event in a small window  $[t, t+\Delta t)$ given the history $\Hcal_t$, \ie,
\begin{align}
  \label{eq:intensity}
  \lambda(t|\Hcal_t) \, \Delta t := \PP\cbr{\text{event in $[t, t+dt)$}|\Hcal_t} = \EE[dN(t) | \Hcal(t)], 
\end{align}
where $dN(t) = N(t+\Delta t) - N(t)$ and one typically assumes that only one event can happen in a small window of size $\Delta t$,
$\ie$, $dN(t) \in \cbr{0,1}$.
 From now on, for the sake of convenience, we will not explicitly denote the dependence on past history, $\mathcal{H}_t$. 
 Furthermore, we can express the log-likelihood of a list of events $\cbr{t_1,t_2,\ldots,t_n}$ in an observation window $[0, T)$ as 
\begin{align}
  \label{eq:loglikehood_fun}
  \Lcal = \sum_{i=1}^n \log \lambda(t_i) - \int_{0}^T \lambda(\tau)\, d\tau,
\end{align}
This simple log-likelihood will later enable us to learn the parameters of our model from observed data.

The functional form of the intensity $\lambda(t)$ is often designed to capture the phenomena of interests. For instance, in a {\emph homogeneous Poisson process}, the intensity is assumed to be independent of the history $\Tcal$ and constant over time,~\ie, $\lambda(t) = \lambda_0 \geqslant 0$. In an {\emph inhomogeneous Poisson process}, the intensity is also assumed to be independent of the history $\Tcal$ but it can be a function varying over time,~\ie, $\lambda(t) = g(t) \geqslant 0$. 

A one-dimensional Hawkes process \cite{hawkes1971spectra} is:
\begin{equation}
\lambda(t) = \mu(t) + \int_{-\infty}^t \kappa(t-s) dN(s)
\end{equation}
Here, $\mu(t)$ is the base intensity, and $\kappa$ is the decaying kernel. The process is well-known for its self-exciting nature; each time a new event happens, the intensity grows by $\kappa(0)$. Then, as time passes, it decreases exponentially to $\mu(t)$.  Hawkes process also exhibits a bursty nature, hence it is also referred to as time-cluster process \cite{hawkes1974cluster}.

In $U$-dimensional Hawkes process, there are $U$ processes that are coupled with each other, and the interactions between processes are explicitly modeled. Formally, $\{N_u(t) | u=1, \ldots, U\}$ is a $U$-dimensional Hawkes process where the conditional intensity for the $u$-th dimension is given by:
\begin{equation}
\lambda_u(t) = \mu_u(t) + \sum_{u'=1}^U \int_{-\infty}^t \kappa_{uu'}(t-s)dN_{u'}(s)
\label{condint}
\end{equation}
where the summation captures the mutual excitation of the processes; i.e., the effect of the occurrence of events in one dimension on the likelihood of future events in all dimensions.
\label{ss_bg}

\vspace{-4mm}
\section{Proposed Method}
\vspace{-2mm}

In this section we formally define the newsfeed prioritization problem, and present our point process model along with the feature that constitute the parameters of our model.

Here, \emph{post} and \emph{comment} can be regarded as an abstraction of any activity in social networks such as likes, comments, photo or status updates, etc. They are intended to show the interaction between activities. One can easily extend the model to capture different types of actions in social networks.


\vspace{-2mm}
\subsection{Problem Definition}
\vspace{-2mm}
In the newsfeed prioritization problem, we are given $C$ cascades and intend to sort them in order to show them on user $u$'s news feed in a way that more attractive cascades to the user end up at the top of his newsfeed.
Each cascade $c$ is represented as a set of events $c = \cbr{e_0, e_1, \ldots, e_{n_c}}$ where $e_0$ indicates the post that initiates the cascade, and $e_1, e_2, \ldots, e_{n_c}$ are the $n_c$ comments this post has received so far.
Each event (either post or comment) $e_i$ is represented as a triple $e_i = (t_i, p_i, d_i)$ where $t_i$, $p_i$, and $d_i$ represent the time of the event, its publisher, and its content, respectively. 

As explained in section \ref{ss_bg}, the conditional intensity function \ref{condint} can be interpreted as the expected rate of event occurrence. A process with higher intensity at time $t$ is more likely to bring about an event than a process with less intensity. Given that no two events occur at the exact same time, the intensity function will be proportional to the probability of the event occurrence. Therefore we use equation \ref{condint} as a measure of user's contribution likelihood to a cascade. This explains why we propose to personalize users newsfeeds according to the value of the intensity function.

\vspace{-2mm}
\subsection{Point Process Model}
\vspace{-2mm}

In this section, we propose a point process model designed to capture the properties we listed in section \ref{motivation}.
Our goal is to find the $post$ and $comment$ $influence$ parameters, and use them to find the likelihood of target user $u$'s contribution to each of the $C$  posts, and finally, sort $u$'s newsfeed.

The likelihood user $u$ comments on cascade $c$ is given by:
\begin{align*}
\lambda_{uc}(t) = \underbrace{\mu_{u e_0} \exp(-\omega_{\mu} (t-t_0))}_{\text{post influence}} + \underbrace{\sum_{t_i < t} a_{u e_i} \exp(-\omega_a (t-t_i))}_{\text{comment influence}}
\end{align*}
where $\mu_{ue_0}$ is the initial influence of post $e_0$ on user $u$, and $a_{u e_i}$ is the initial influence of comment $e_i$ on user $u$ for $i \ge 1$. As time goes on, these influences decrease exponentially with rates $\omega_\mu$ and $\omega_a$, respectively.

As a reminder of our discussion in section \ref{motivation}, the influence of an event on user $u$ is a combination of the influence of the event publisher and the influence of the event content influence. As a result, we rewrite post and comment influences as follows:
\begin{align}
& \mu_{u e_0} = \mu_{u p_0} + \mu_{u d_0} \\
& a_{u e_i} = a_{u p_i} + a_{u d_i}
\end{align}
where the subscript $up_i$ indicates the influence of the event publisher $p_i$ on user $u$, and  $ud_i$ indicates the attractiveness of the event content $d_i$ over $u$.
Table \ref{tb:model} provides a summery of notations used in this subsection. Moreover,
Hawkes process, as a temporal point process, naturally respects the rest of the properties listed in section \ref{motivation}, namely \textbf{time subordination}, \textbf{self excitation}, \textbf{mutual excitation}, and \textbf{bursty nature} of events.
\begin{table}[t]
\centering
\caption{Dynamics of conversations and the corresponding parameters of the proposed model}
\label{tb:model}
 \rowcolors{2}{white!25}{gray!25}
\begin{tabular}{|c|c|} \hline
 \rowcolor{gray!75}
 Observed dynamics & Proposed model \\ \hline
  Post Publisher Influence & $\mu_{up_0}$ \\\hline
  Post Content Influence & $\mu_{ud_0}$  \\ \hline
  Post Influence Decay & $\exp(-\omega_{\mu} \Delta t)$ \\\hline
  Comment Publisher Influence & $a_{up_i}$ \\ \hline
  Comment Content Influence & $a_{ud_i}$  \\ \hline
  Comment Influence Decay & $\exp(-\omega_{a} \Delta t)$ \\ \hline  
   \end{tabular}

\label{table1}
\vspace{-3mm}
\end{table}

Up to now, the number of parameters to be learned is at least $O(U^2)$. Given that the fact that there is not always as much training data available as required, it is practically impossible to learn the model effectively.
To address this problem, we introduce a set of features to derive the influence between users via a weighted sum of them. These features include the social profiles of both sides of an event, their interaction history, and linguistic and psychological features of the content of the event. The following is the feature-based representation of the model parameters:
\begin{align}
& 
\mu_{u p_0} = \alpha^{\top}  \Fcal_{u p_0}
&
 \mu_{u d_0} =  \beta^{\top}  \Fcal_{d_0}  \\
& 
a_{u p_i} = \gamma^{\top}  \Fcal_{u p_i} 
&
a_{u d_i} = \sigma^{\top}  \Fcal_{d_i}
\end{align}
Here, $\Fcal_{u p}$ indicates the features extracted from the interaction history of $u$ and $p$, and $\Fcal_{d}$ indicates the content features. Vectors 
$\alpha$, $\beta$, $\gamma$, and $\sigma$ are the feature coefficients, or weights, and are the only parameters to be learned. They act as a measure of the importance of each feature in building up the likelihood.
As you see, our learning space is reduced to the number of feature coefficients, $O(K)$ where $K$ is usually much smaller than $U$.
In the experiments section we show how this meaningful reduction of the parameters leads to better results.

\subsection{Efficient Parameter Estimation}
\vspace{-2mm}
Assuming we have previously observed $C$ cascades first we need to learn the parameters and then use them in prioritization task. 
Without loss of generality we assume all posts arrive at time $t_0 = 0$ to write a simpler form for likelihood function.
Given the parameters, the cascades are independent, therefore, the log-likelihood can be written as
$\Lcal = \sum_{c=1}^C \Lcal_c$, where, 
\begin{align*}
\Lcal_c = 
&  
\sum_{i=1}^{n_c} \log \lambda_{p_i c}(t_i) - \sum_{u=1}^{U} \int_0^{T} \lambda_{uc}(t) dt   \\ 
= &  \sum_{i=1}^{n_c} 
 \log \Bigl( (\alpha^{\top} \Fcal_{p_i p_0}+\beta^{\top} \Fcal_{d_0}) e^{-\omega_\mu (t_i-t_0)} \\
 & \hspace{1.2cm} +  \sum_{j < i}  (\gamma^{\top} \Fcal_{p_i p_j}+\sigma^{\top} \Fcal_{d_j}) e^{-\omega_a (t_i-t_j)}  \Bigr) \\
& 
-\sum_{u=1}^{U}  (\alpha^{\top} \Fcal_{p_i p_0}+\beta^{\top} \Fcal_{d_0}) (1- e^{-\omega_\mu (t_i-t_0)})/\omega_\mu \\
&  
- \sum_{u=1}^{U} \sum_{j=1}^{n_c} (\gamma^{\top} \Fcal_{u p_j}+\sigma^{\top} \Fcal_{d_j}) (1-e^{-\omega_a (t_i-t_j)})/\omega_a 
 \end{align*}
 
Perhaps surprisingly, the log-likelihood is convex with respect to parameters $\alpha$, $\beta$, $\gamma$, and $\sigma$, therefore these parameters can be estimated using any of the well-developed convex optimization method, maximizing the log-likelihood under the constraint of being element-wise positive.

To preserve convexity while prohibiting over-fitting, we use $l_1$ norm to regularize the model.
\begin{align}
\min_{\alpha, \beta, \gamma, \sigma} - \Lcal + \zeta_\alpha  ||\alpha ||_1 + \zeta_\beta ||\beta ||_1
+ \zeta_\gamma  ||\gamma ||_1 + \zeta_\sigma ||\sigma ||_1
\end{align}
subject to $\alpha_u \geq 0$, $\beta_u \geq 0$, $\gamma_u \geq  0$, and $\sigma_u \geq  0$ for all $u$ where $\zeta_\alpha$,  $\zeta_\beta$,  $\zeta_\gamma$,  $\zeta_\sigma$ are regularization parameters. 

\vspace{-2mm}
\subsection{Prioritization}
\vspace{-2mm}

Having learned the model parameters, we can now sort the events on the news feed of a target user $u$ based on $\lambda_{uc}(t)$ for all cascades $c$ at every time $t$. According to equation \ref{eq:intensity} intensity is the average rate of commenting or equivalently proportional to the probability of commenting given the history. $\ie$,
\begin{align}
  \label{eq:intensity}
  \lambda(t) \propto \PP\cbr{\text{event in $[t, t+dt)$}|\Hcal_t} 
\end{align}
and that's why it is very suitable candidate to serve as a prioritization measure.

It is notable that we do not need to compute everything from scratch when a new event occurs. The summation on previous events can be updated on the fly from the quantity at the previous time thanks to ideas borrowed from dynamic programming.
To elaborate, let us assume we have computed $\lambda_{uc}(t)$ at some time $t_1$ as the summation of two terms, \ie, $\lambda_{uc}(t) = A(t) + B(t)$ where
$A(t) = 
\mu_{u e_0} \exp(-\omega_{\mu} (t-t_0))
$ and 
$B(t) =
\sum_{t_i < t} a_{u e_i} \exp(-\omega_a (t-t_i))
$.
It can be easily seen that the intensity at some time $t_2 > t_1$ can be easily updated by setting $A(t_2) = A(t_1) e^{-\omega_\mu(t_2-t_1)}$, and $B(t_2) = B(t_1) e^{-\omega_a(t_2-t_1)}$.

\subsection{Features}
\vspace{-2mm}

The proposed framework is general enough to work with any feature extracted from the users (or even without any feature in its simplest form), such as their profile information and interaction history. In this subsection, we briefly introduce the features we use to parametrize the influence. Remember we have different influence parameters for posts and comments, namely, \emph{publisher influence} and \emph{content influence}.
We believe that the social influence of the publisher on the user is a function of the social status of both, as well as their history of interactions. Therefore, \emph{Character-based influence} has to do with characteristics of the publisher and the user, for example, their popularity, knowledge, or level of activity; On the other hand, \emph{relationship-based influence} is a measure of the history of interactions between the publisher and the user, such as shared interests and activities. Table \ref{tb:features}
 demonstrates the features we extracted for each of these two elements of influence. Note that we extract Relationship based features for the publisher and the users separately to respect the directed nature of social interactions. 

Similarly, we divide the content influence into two elements: linguistic features, and emotional/psychological characteristics extracted from the content. 

We make use of Linguistic Inquiry and Word Count (LIWC) to extract content features \cite{pennebaker2001linguistic}. LIWC is a text analysis software that calculates the degree any text comprises different categories of words, and is used in several prior studies in conversation analysis \cite{budak2013participation, ireland2011language}. 

As given in Table 2, to parametrize the influence of the \emph{publisher} on the \emph{user} in a post/comment, we use 35 features categorized into the following sets:

\textbf{ChrPub:} Character-based features of publisher

\textbf{ChrUser:} Character-based features of user

\textbf{RltnPub:} Relationship-based features of publisher

\textbf{RltnUser:} Relationship-based features of user

\textbf{Lng:} Linguistic features of the post/comment 

\textbf{Psy:} Psychological features of the post/comment


\begin{table}[t]
\centering
\caption{Features to parametrize the influence}
\label{tb:features}
\begin{tabular}{|l|l|l| }
\hline
Class & Feature & Explanation/Example \\ \hline \hline
\multirow{6}{*}{\rotatebox[origin=c]{75}{Character} } 
&  $\diamond$ Activity & $\#$ posts one makes  \\
&  $\diamond$ Attractiveness &  \# comments once posts receive \\
&  $\diamond$ Sociability & \# comment one makes \\
&  $\diamond$ Responsiveness & \# posts one comments on \\
&  $\diamond$ Connectivity & \# users one ever \\
& &  commented on their posts \\
\hline
\multirow{10}{*}{\rotatebox[origin=c]{75}{Relationship}  } 
& $\diamond$ Post  & \# comments one makes \\ 
& Influence  & on the other's posts  \\
& $\diamond$ Comment &  \# comments one makes \\
& Influence  &  after the other's comment \\
& $\diamond$ Direct post  & \# times one is the first  \\
& Influence & commenter on the other's post \\
& $\diamond$ Direct comment & \# times one is the first  \\ 
& Influence   &commenter on the other's post \\
&  $\diamond$ Co-commenting & \# posts one comments after \\
& & the other's comment \\
\hline
\multirow{8}{*}{\rotatebox[origin=c]{75}{Language}  } 
&  $\diamond$ Word count & - \\
 &  $\diamond$ Words>6 letters & - \\
 &  $\diamond$Total pronouns & I, them, itself  \\
 &  $\diamond$ Common verbs  & Walk, went, see  \\
 &   $\diamond$ Adverbs   & Very, really, quickly  \\
&   $\diamond$ Quantifiers & Few, many, much  \\
 &   $\diamond$ Numbers  & Second, thousand  \\
 &   $\diamond$ Swear words  & Damn, piss, fuck  \\ \hline
\multirow{7}{*}{\rotatebox[origin=c]{75}{Psychology}  } 
&  $\diamond$ Social processes & Mate, talk, they, child  \\ 
&  $\diamond$ Affective processes & Happy, cried, abandon  \\
&  $\diamond$ Positive emotion & Love, nice, sweet   \\ 
&  $\diamond$ Negative emotion & Hurt, ugly, nasty   \\
&  $\diamond$ Cognitive processes  & Cause, know, ought  \\
&  $\diamond$ Perceptual processes  & Observing, heard, feeling  \\
&  $\diamond$ Biological processes  & Eat, blood, pain  \\
\hline
\end{tabular}
\vspace{-5mm}
\end{table}

\vspace{-1mm}
\section{Related Works}
\vspace{-1mm}
The problem of personalizing news feeds in online social media has been recently addressed in several ways. The approaches taken to this problem can be categorized by the ranking criteria they use:

\emph{Global criteria (Publisher authority and global content attractiveness):} Gabrilovich \ea~\cite{gabrilovich2004newsjunkie} propose a framework for comparing text documents using language models, and sort users' newsfeed based on the novelty and relevance of the content items. Shmueli \ea~\cite{shmueli2012care} proposed a content-based collaborative filtering method based on co-commenting and textual tags to predict which stories a particular user is most likely to comment on. Yi \ea~\cite{yi2014beyond} integrate dwell times of users on content items into collaborative filtering and machine learning-to-rank models.

\emph{Relationship-based criteria (User-to-user relationship and content relevance):} Authors in~\cite{backstrom2013characterizing} pose the re-entry prediction problem for the first time and extract several social, temporal, and textual features of the post, poster, comments and commenter to address it. Authors in~\cite{chen2012collaborative} also utilize a comprehensive collection of social, textual and content relevance features along with publisher authority in a collaborative filtering approach to Tweet recommendation. In~\cite{uysal2011user}, a similar set of features is used to filter tweets based on users' retweet likelihood. In ~\cite{freyne2010social}, the authors measure the significance of a similar set of features in ranking feeds of IBM's SocialBlue and show that browsing history is a more accurate predictor of content relevance than communication history.

Phelan \ea \cite{phelan2009using} use Tweets to propose a content-based recommendation system to rank news stories in RSS feeds by calculating TF-IDF scores to measure the co-occurrence of popular terms within the user's RSS and Twitter feeds.

Inspired by physical interactions, authors in~\cite{budak2013participation} study the factors affecting user participation in the context of Twitter chats, however, their work is limited to the re-entry problem.

Our work is similar to the above mentioned methods in that it also seeks to rank a set of content items based on a personalized measure of relevance. But this paper differs from this direction of related work in that these methods disregards many unique properties of social media such as publisher influence and self- and mutual-excitation, as well as the bursty behavior of individuals and the evolution of dynamics with time. We use a comprehensive set of global and user-specific features to model conversation dynamic and predict events.

To the best of out knowledge, no recent work has employed temporal point processes (or Hawkes Process) to model conversation dynamics on social networks. Our model captures self and mutual excitation of user behavior to predict events based on user profile, interaction history, and content popularity.

The use of Hawkes process has been reported in the study of the association of temporal events in various fields, for example, financial events~\cite{embrechts2011multivariate}, seismic events~\cite{marsan2008extending}, crimes~\cite{stomakhin2011reconstruction},
civilian deaths in conflicts~\cite{lewis2011self}, and
recently, causal militant conflict events~\cite{li2013dyadic}, social media~\cite{zhou2013learning,yang2013mixture,farajtabar2014shaping}, and network evolution modeling~\cite{FarWanGomLiZhaSon15}.
Authors in~\cite{wang2012user}  utilize inhomogeneous poisson processes to deal with the item adoption problem.
Zhou \ea~\cite{zhou2013learning}, exploit multi-dimensional Hawkes process with low-rank and sparse assumption to accurately infer the influence network of WWW merely via the timestamps of new links between websites.
In~\cite{farajtabar2014shaping}, authors proposed a novel framework based on Hawkes process to capture the dynamics of product adoption in social networks. They introduced a variety of activity shaping problems that which are convex and scalable and can be used to actively manage social networks.
Exact/approximate (non) parametric estimation algorithms for Hawkes process are also proposed by various
authors~\cite{olson2013exact,zhou2013learning,zhou2013kernel}. 
These methods suffer from quadratic number of parameters and an inherent batch learning procedure.

\vspace{-1mm}
\section{Experiments}
\vspace{-1mm}
\subsection{Dataset}

\begin{table*}[t]
\centering
\caption{Dataset}
\label{tb:dataset}
 \rowcolors{2}{white!25}{gray!25}
 
\begin{tabular}{|c c c c c c|} \hline
\rowcolor{gray!75}
ID & FB Handle & Name and Brief Introduction &  \# Users & \# Posts & \# Comments \\ \hline \hline
1 & 151761649081 & Dystopia Rising- A community of gamers & 1100 & 2600 & 68140\\ \hline
2 & 244137892388551 & Paul's Icmeler Lovers- Fans of Icemeler town in Turkey & 1599 & 1702 & 15968\\ \hline
3 & 314680351956164 & A discussion group about Scotland's independence & 2488 & 6481 & 129458\\ \hline
4 & 251598898325733 & Bowling Arguments & 521 & 589 & 10511 \\ \hline
5 & 221937001327409 & Relationship Talk 2.0 & 1123 & 5922 & 128646 \\ \hline
6 & 2539771528 & Debate- An open discussion group & 1860 & 3334 & 165062\\ \hline
7 & 422219507812142 & Kwinana Chat- Residents of a town in western Australia & 2658 & 7571 & 76964\\ \hline 
8 & 1401143476785413  & VetChat- A community of horse keepers & 1184 & 752 & 9025\\ \hline
9 & 383886871680383 & An open discussion group & 93 & 427 & 11698\\ \hline
10 & 20444826822 & Beyond Cesar Millan- A community of dog owners& 1542 & 1210 & 23159\\ \hline
11 & 156560987806631 & Bridgnorth chat, news, rants and idle speculation & 2853 & 5923 & 60212 \\ \hline
12 & 112895548793967 & A community of Obama supporters& 1247 & 1784 & 9545 \\ \hline
13 & 571563396239160 & Debate- Another open discussion group & 465  & 360 & 8357 \\ \hline
14 & 488292291204196 & Does God Exist? & 2310 & 5782 & 238498 \\ \hline
15 & 114421575256439 & RationalWiki- A discussion group against pseudoscience  & 1059 & 4293 & 51261 \\ \hline
16 & 343934742322479 & Scottish Secular Society & 526 & 10042 & 71734 \\ \hline
\end{tabular}
\end{table*}

We examine our model on discussion cascades crawled from Facebook groups. Facebook groups are online communities of people who share interests or concerns, such as coworkers, classmates, celebrity fans, parents, etc. People join Facebook groups to share updates, photos and documents and receive feedback on them.  With hundreds of millions of online groups each with tens to thousands of members discussing topics ranging from pet care to philosophy, Facebook provides us with abundant ongoing conversations among a known set of people to alleviate the study of conversation dynamics in online social media. 
We have collected a corpus of around 50,000 conversations recently drawn from Facebook, with over 1 million comments involving about 25,000 people. To our knowledge, this is the only corpus of Facebook groups conversation data available for study so far.
The groups selected for this study were 16 open groups (groups whose posts can be viewed by non-members) chosen to cover a range of topics and populations presented in Table \ref{tb:dataset}.
From each group, we collected as many conversations from the group's newsfeed as possible for our headless browser. The resulting corpus consists of 1078238 comments, naturally partitioned into 58772 posts among 22628 users. 

It is notable that there is a minor bias in our dataset. The training data is collected from Facebook groups feeds. Since Facebook applies its own ranking of events in group news feeds, the collected data is biased toward attractive events. However, we argue that this bias has no significance in model evaluation, as both the training and test datasets are affected by it. A model learned from a set of users in one OSM, say Facebook, may not work properly on test data from the same set of users on another OSM, say Twitter. This is minor problem as the models are usually trained and utilized on the same environment. A discussion of the effect of this bias on the descriptive power of our model is given in section \ref{sec:pred}.

\subsection{Prioritization}

In this section, we test the proposed method and compare its effectiveness to a number of baselines each having a special property that makes them interesting to discuss.
\subsubsection{Experimental Setup}

\textbf{Pre-processing and initialization}
For each group, cascades are sorted according to their initiation times. Depending on the size of the group $C$, a fraction of cascades is used as training data. Test posts are then chosen from the remaining posts whose  participants are among the participants in the training data. Table \ref{tb:prior} gives an overview of the data available after this preprocessing step. Time is measured in minutes. We set $\omega_{\mu}=0.001$ and $\omega_a=0.01$ to reflect the decay rates of the post and comment influence, respectively. This choice of decay rates means that a post/comment will loose 65 \% of their influence after 1000/100 minutes and is estimated according to the conversation progress rate of the group at study. The regularization parameter $\alpha$ is set by cross validation to the best performance. The features are  extracted from training data and are normalized to lie within $[0,1]$. 

\textbf{Evaluation Metric.} After we train our model on training data, we proceed with the test data one comment at a time, and sort the posts for every user using the  algorithms in study. We assign a rank to the comment at hand, with regard to the rank of the post it concerns. For example, if the newly arrived comment concerns the top post in our sorted set, we assign 0 to the comment. If the comment concerns the second top post, we assign 1, and so on.  Finally, we compute the average rank over all  incoming comments. This equips us with a measure called \verb+AveRank+ of the algorithm, that, intuitively, is the average number of posts each user is bound to scroll down to find their topic of interest when their newsfeed is sorted according to the algorithm is study.

Note that \verb+AveRank+ is a means of intra-group performance analysis. However, since some groups are more active and the user faces many more candidates when deciding to contribute to threads in these groups than others, it is essential to define a measure that enables us to conduct inter-group analysis.  We introduce the \emph{normalized average rank}:
\begin{equation}
NAveRank = AveRank  /  \overline {Activity}_g
\end{equation}

where $\overline{Activity}_g$ of group $g$ is its number of active cascades averaged over time. We define active cascades as those that received their last comment no more than 12 hours ago.

\textbf{Methods.} We compare the proposed method to the following four methods:

 \verb+Reverse Chronological baseline (RCHR)+. This method sorts cascades according to their most recent timestamp, whether it is the initiation timestamp or the last comment timestamp). There are still many online social networks and communities that use this baseline criterion to sort news feeds.

\verb+Nearest Neighbor (NN)+. In this baseline method, for each user, we compute the moving average of feature vectors of their past comments. We then rank the given posts according to their distances from the moving average. This method uses all 35 features.

 \verb+Cox process (COX)+. In this method, we implement a Cox process in which the likelihoods of future comments on each cascade are used to rank the user's interests in the thread. The likelihood
of an incoming comment by user $u$ on cascade $c$  is:
$
\lambda_{uc}(t) = \exp\{\boldsymbol{\rho}^T \Fcal_d(t)\}
$.
The weight vector  $\boldsymbol{\rho}$ can be estimated via the maximization of the \textit{Partial Likelihood} \cite{perry2013point} using any convex optimization method. Since character and relationship based features are not usable with Cox process we used linguistic and psychological feature sets to come up with two algorithms; \verb+COX-LNG+ and \verb+COX-PSY+. This method represents temporal point processes lacking mutual excitation.

\verb+Hawkes (HWK)+. This method is the basic Hawkes process before incorporating features. The likelihood that cascade $c$ receives a comment from user $u$ is based only on past events, and the social influence of the post publisher and  commenters on $u$, and is given by:
\begin{equation*}
\lambda_{u}(t) = \mu_{up}g(t-t_0) + \sum_{t_0 < t_i < t} a_{up_i} g(t-t_i),
\end{equation*}
where $\mu_{up}$ is the influence of the post publisher $p$ on $u$, and $a_{up_i}$ is the influence of the commenter  $p_i$ on $u$. The parameters  $\boldsymbol{\mu}$ and $\mathbf{a}$
can be estimated efficiently using  EM \cite{zhou2013learning}. This algorithm is implemented to show how parameterizing the influence benefits in cases of data scarcity.

 \verb+Feature-based Hawkes (FHWK)+. This is the method proposed in the current paper. We have utilized five different feature sets to introduce five versions to the method: character-based (ChrPub and ChrUser features), relationship-based (RltnPub and RltnUser features), linguistic (Lng features), and psychological (Psy features) versions, respectively referred to as  \verb+HWK-CHR+,  \verb+HWK-RLTN+,  \verb+HWK-LNG+, and  \verb+HWK-PSY+.  We introduce the fifth version by utilizing all the four feature sets \verb+HWK-ALL+.

\subsubsection{Analysis}

\begin{figure}[t]
\centering
\includegraphics[width=.39\textwidth]{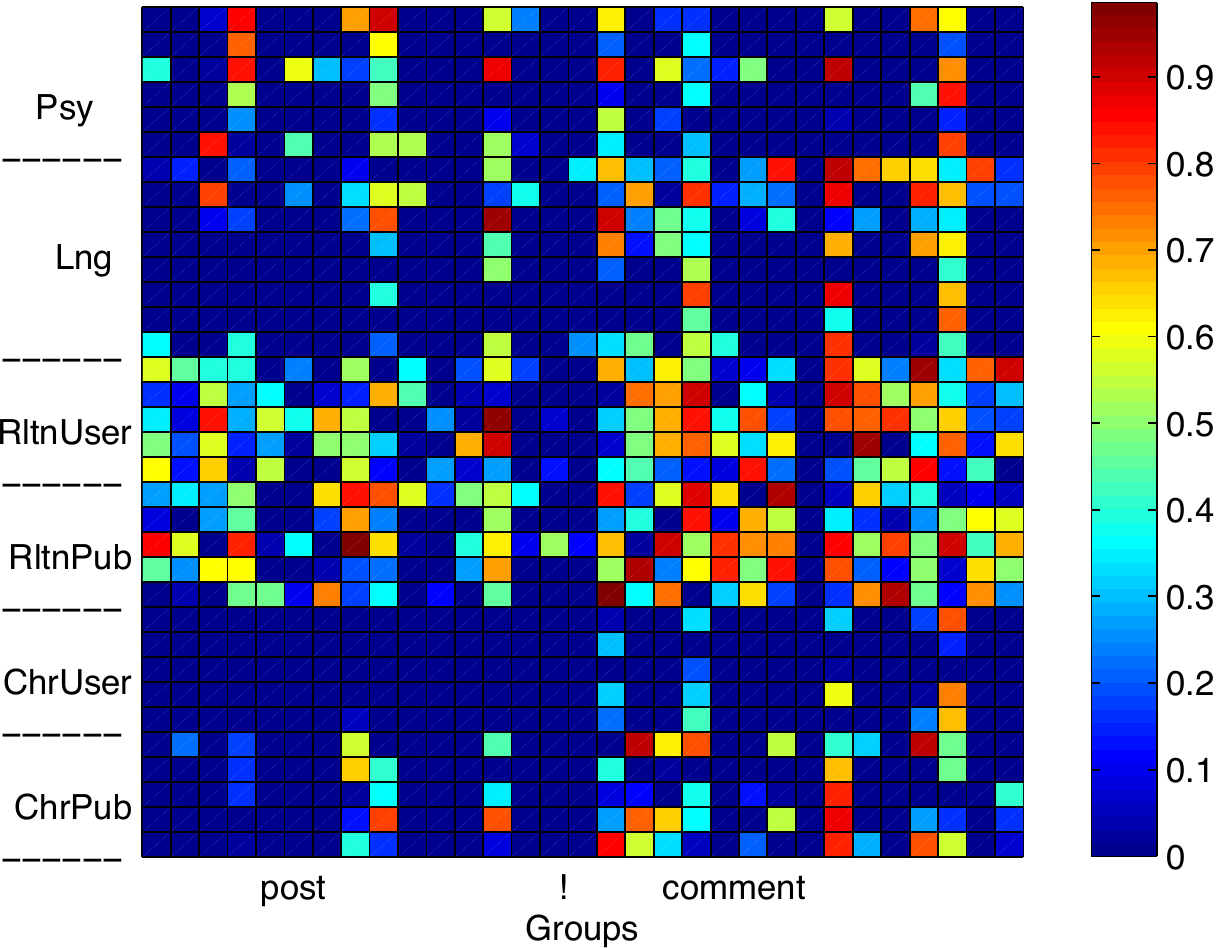}
\caption{Post and comment influence parameters for all groups. All 35 features are utilized. }
\label{group-feature-all}
\end{figure}

Let's start with a visual analysis of the proposed method. Figure \ref{group-feature-all} shows the 35 learned parameters for all the datasets via \verb+HWK-ALL+.  One can evaluate the significance of each of the 35 features in comprising the influence using this figure. From \ref{group-feature-all}, the following observations can be made:
\begin{itemize}
\item[-] The columns are mostly sparse as imposed by regularization. About half of the elements are close to 0.
\item[-] A great weight is concentrated on the relationship-based features, as in accordance with our expectation. These features correlate very well with the observed interactions in the training data. However, as we will see later, they do not guarantee  high predictive ability. 
\item[-] There is a notable emphasis on the Publisher parameters compared to User parameters, revealing that the characteristics of the publisher are more influential than those of the user. In contrast, for relationship-based features, it seems that both the publisher and the user contribute equally in the influence, suggesting a reciprocity in the group's interactions.
\item[-] It seems that different groups share the same dynamics. Although they have different weights for some of the features, the majority of the features play a similar role in all groups.
\item[-] The post and comment parameters  within individual groups do not take on the same values, but are highly correlated. It implies that the same features derive post and comment influences, but not with the same strength, which is merely a difference in scale.
\item[-] In small and low activity groups, influence is more dependent on characteristic features of publisher and user than on the rest of the features.
\end{itemize}
 
\begin{figure} [t]
        \centering
        \begin{subfigure}[b]{0.33\textwidth}
                \includegraphics[width=\textwidth]{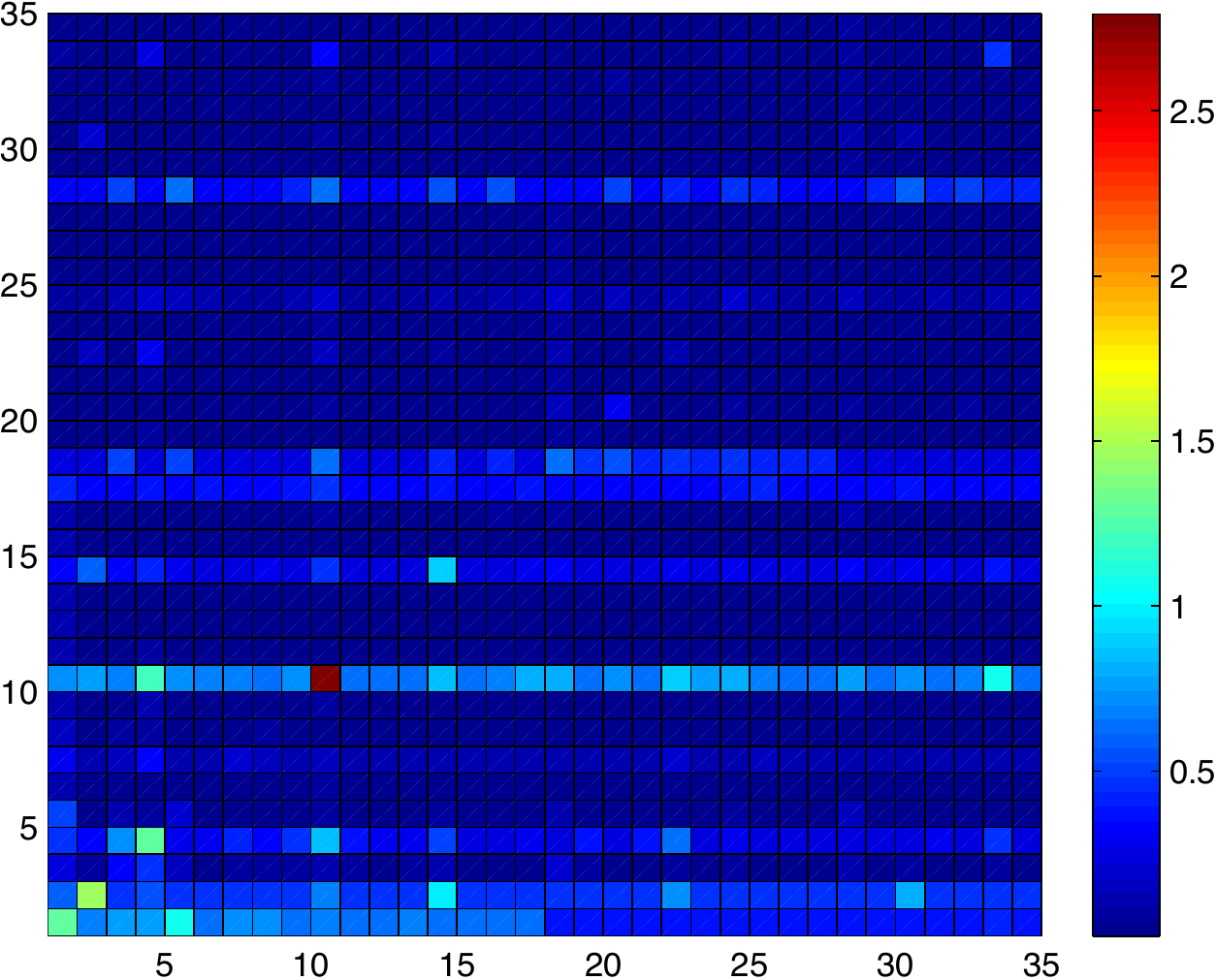}
                \caption{Post influence, $\boldsymbol{\mu}$}
                \label{fig:gull}
        \end{subfigure}%
        
        \begin{subfigure}[b]{0.33\textwidth}
                \includegraphics[width=\textwidth]{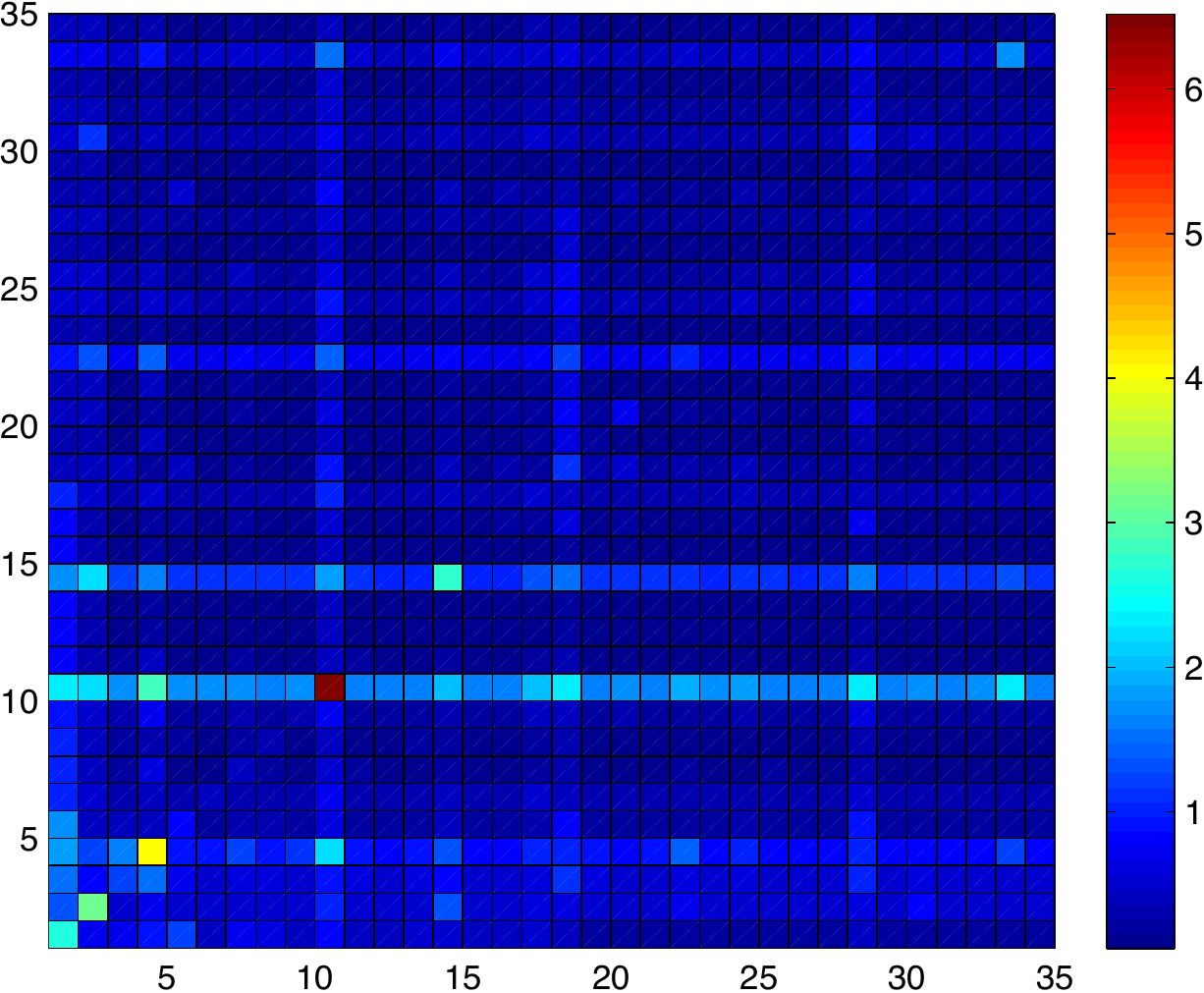}
                \caption{Comment influence, $\boldsymbol{a}$}
                \label{fig:tiger}
        \end{subfigure}
        \caption{Mutual influence between first 35 users in group 3}
        \label{fig:inf}
\end{figure}

Figure \ref{fig:inf} demonstrates the directed pairwise influence among users in one of the groups. The influence is obtained by the dot product of the features and the corresponding coefficients learned by \verb+HWK-ALL+.
For clarity, we included the first 35 users only. Note the following observations:

\begin{itemize}

\item[-] Influence matrix is sparse as in accordance with our intuition of real-world networks, where each individual is connected to and influenced by a limited number of people only.
\item[-] A correlation between post influence and comment influence is observed. If $p$'s posts are attractive to $u$, so are her comments. 
\item[-] The influence of user 10 on others (column) and the influence she receives from others (row) are both remarkable. Examining the dataset reveals that she is a super-active user but she also is a late adopter: she usually comments after a post receives a sufficient number of comments. This explains why her input influence (row) is greater than her output influence (column).
\item[-] Diagonal values of the matrix seem to mostly take on large values. These parameters reflect the self-excitation of individual activities. If a person comments on a post, it is likely that they return to the post to recontribute.
\item[-] Off-diagonal values such as the square in row 5 and column 10 in both matrices, indicate notable mutual excitation.
\end{itemize}

\begin{table*}[t]
\centering
\scriptsize
\caption{Prioritization results for all groups and algorithms in terms of normalized average rank}
\label{tb:prior}
\rowcolors{2}{white!25}{gray!25}
\begin{tabular}{| c || c c c || c c c c c c c c c c |} \hline 
 	\rowcolor{gray!75}
 ID & \# train posts& \# test posts & \# users & \verb+RCHR+ & \verb+NN+ & \verb+COX-LNG+ & \verb+COX-PSY+ & \verb+HWK+ &   \verb+HWK-CHR+ &   \verb+HWK-RLTN+ & \verb+HWK-LNG+ &  \verb+HWK-PSY+ &  \verb+HWK-ALL+  \\[0pt] \hline \hline
1 & 	500	& 675 & 243&	7.19 & 3.35 & 3.68 & 3.53 & 3.12 & 0.90 & 1.20 & 1.14 & 1.54 & 1.05   \\[0pt] \hline
2 &600 &	460 & 872&	1.16 & 0.82 & 1.99 & 2.18 & 5.00 & 0.29 & 0.51 & 0.30 & 0.36 & 0.40   \\[0pt] \hline
3 & 500 &	 500 & 392&	3.98 & 2.63 & 2.95 & 4.08 & 4.60 & 0.88 & 1.51 & 1.29 & 1.21 & 1.20  \\[0pt] \hline
4 & 200 &	52 	& 186& 	0.51 & 0.87 & 5.07 & 5.25 & 3.67 & 0.45 & 0.57 & 0.56 & 0.56 & 0.51  \\[0pt] \hline
5 &1000 &1000	 & 514 & 	0.22 & 0.14 & 0.30 & 0.46 & 0.21 & 0.14 & 0.19 & 0.29 & 0.10 & 0.20 \\[0pt] \hline
6 & 500 &	337	& 558 & 	1.44 & 0.36 & 4.66 & 7.54 & 2.47 & 0.38 & 0.39 & 0.50 & 0.42 & 0.38  \\[0pt] \hline
7 &800 &	1000	 &  950&	4.83 & 4.86 & 5.31 & 6.23 & 13.09 & 0.52 & 0.60 & 0.53 & 0.62 & 0.60  \\[0pt] \hline
8 &400 &	73 &	719		& 2.14 & 2.40 & 8.86 & 9.97 & 56.04 & 0.91 & 0.92 & 0.92 & 0.97 & 0.91  \\[0pt] \hline
9 & 200 &	110 &	64 	& 1.34 & 0.63 & 3.06 & 2.48 & 2.18 & 0.62 & 0.64 & 0.61 & 0.62 & 0.61 \\[0pt] \hline
10 & 600 &223 &979	& 1.23 & 1.77 & 5.50 & 6.43 & 9.20 & 0.34 & 0.48 & 0.38 & 0.45 & 0.39 \\[0pt] \hline
11 &2000 &1950 &1791	&7.84 & 5.87 & 3.88 & 5.55 & 15.60 & 0.56 & 0.58 & 0.55 & 0.59 & 0.54  \\[0pt] \hline
12 &600 &600	 &764	 & 1.19 & 0.34 & 0.35 & 0.24 & 3.01 & 0.37 & 0.85 & 0.30 & 0.30 & 0.75 \\[0pt] \hline 
13 &70 &	75 &	213		&  0.90 & 1.71 & 12.56 & 11.71 & 11.68 & 0.50 & 0.56 & 0.56 & 0.53 & 0.50  \\[0pt] \hline
14 &1000 &650 & 481 	& 2.49 & 1.15 & 3.58 & 3.64 & 1.16 & 0.38 & 0.45 & 0.46 & 0.62 & 0.44   \\[0pt] \hline
15 &2000	& 1000 &	577 	& 3.79 & 0.95 & 1.95 & 2.40 & 0.44 & 0.62 & 0.71 & 0.68 & 0.61 & 0.82  \\[0pt] \hline
16 &2000 &2000 & 189 	&  4.75 & 0.15 & 0.35 & 0.29 & 0.13 & 0.52 & 0.75 & 0.55 & 0.65 & 0.74  \\[0pt] \hline
avg & & & & 3.71 &	2.16 &	4.46 &	4.93 &	8.61 &	0.63 &	0.83 &	0.74 &	0.82 & 	0.75 \\ \hline 
\end{tabular}
\end{table*}

\subsection{Prediction}
\label{sec:pred}
\begin{figure}[t]
\centering
\includegraphics[width=.42\textwidth]{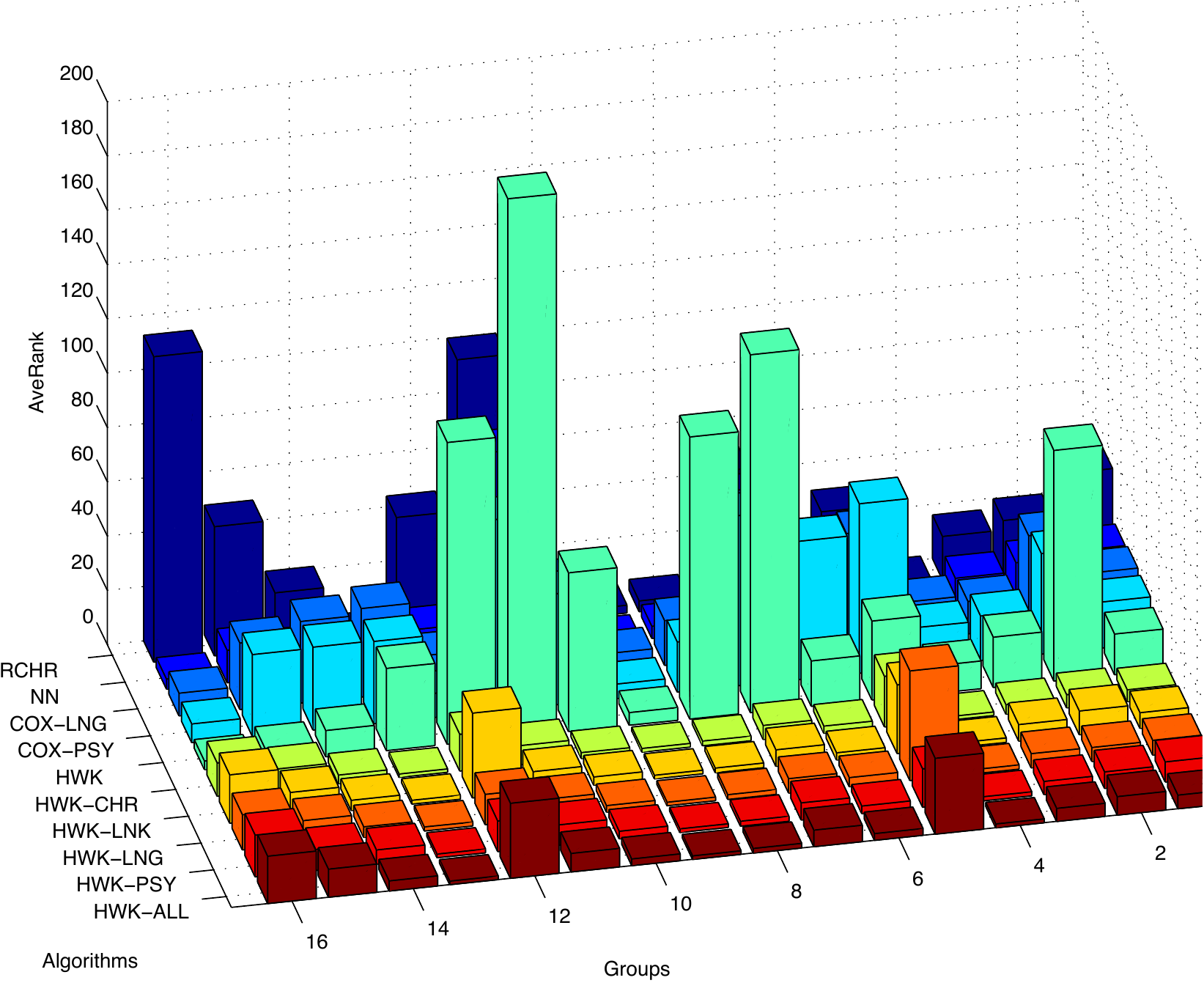}
\caption{Visualization of Table 4. Prioritization results for all groups and algorithms in terms of average rank}
\label{fig:averank}
\end{figure}

We continue by quantitatively evaluating the proposed method. Table \ref{tb:prior} shows \verb+NAveRank+ values obtained using the algorithms listed in subsection 5.2.1 on all 16 groups. To give a better insight into our results, the performance of the algorithms is also represented graphically in Figure \ref{fig:averank} using \verb+AveRank+ measure because of its intuitive comprehensibility.

Note in Table \ref{tb:prior} and Figure \ref{fig:averank} that relationship-based features do not yield good prediction results while being assigned great weights in the model. We think that this is due to the fact that data limitation is part of the problem we are dealing with. To extract descriptive relationship-based features, we need sufficient interaction data. With little data, learning is prone to overfitting and yields poor results on unseen data. On the other hand, our training data is collected from Facebook group newsfeeds, meaning it is biased toward containing cascades with exaggerated cascade re-entry and relationship-based activity. We attribute this to the fact that Facebook sorts each user's newsfeed based on interaction history, and also highly ranks cascades in which the user has previously participated. As a result, our learning algorithm overfits by assigning high values to relationship-based features, features that do not contribute much to the prediction task.

On the contrary, character-based features seem to lead to the best results. Linguistic features are also successful, however, when combined with the rest of the features in \verb+HWK-ALL+, the result deteriorates due to the overlap of correlated features. When two features are correlated, then an increase (or decrease) in the objective function can be attributed to any of them, introducing noise to the model. Another factor contributing to this deterioration is addition of relationship-based features causes overfitting.

In most cases, Hawkes-based algorithms, are superior to all others, and where this is not the case, the difference is very small. We attribute this success to the excitation properties of Hawkes process, and its temporal and bursty nature along with the parameterization of likelihood into features.
Lets examine the behavior of the featureless Hawkes algorithm, \verb+HWK+. In some cases it demonstrates superior performance (groups 15 and 16), and in others quite the contrary (groups 7, 8, 11, 12). Looking into the number of training posts and users of these cases reveals a very interesting result. Superior results are obtained in groups that contain a large number of cascades while maintaining a relatively small number of users. In contrast, inferior performance is obtained on groups with so many users that learning $O(U^2)$ becomes too much.

Normalized average ranking enables further analysis  to proceed from user-level to group-level properties. Consider, for example, the last column of the table \ref{tb:prior} in which results of \verb+HWK-ALL+ is presented. Some groups have higher \verb+NAveRank+  compared to others. The first issue is that these groups are less predictable. A subjective analysis of these groups reveals some fundamental differences in them.

We roughly divide the groups into two contrasting categories: In the first category of groups, socially-oriented conversations take place in which participants know each other to some extent and this acquaintance influences their probability of engaging in a conversation. Groups 2, 4, and 11 fall in this category. The first two are groups of residents of specific areas, and the third is an informal community of bowling players.

The second category consists of groups oriented around specific topics of controversy or specialized discussions. In these groups, whether a user contributes to a thread depends mostly on the topic of the thread, not on the strength of his link with the poster. Participants mostly engage in one-time contributions to threads, although prolonged conversations may also take place. Groups 1 and 8, specialized communities of horse trainers and gamers, respectively, are examples of such groups.
We have noticed that the predictive power of the proposed algorithm is higher when dealing with the first category. As seen in the last column of table \ref{tb:prior}, the normalized average rank of the first category of groups is lower than that of the second category. 

\vspace{-1mm}
\section{ Discussion}
\vspace{-1mm}
Our model can easily be extended to capture different rates of decay for users or groups. Furthermore, we can utilize any engineered feature set or even no feature to learn the Hawkes model. It's notable that \emph{post} and \emph{comment} in this paper can be regarded as an abstraction of any activity in social networks such as likes, comments, photo, status sharing, etc. One can easily generalize the model to capture different types of actions in social networks and model their interaction via coupling their intensities. In our experiments we merely used the usual notion of posts and comments to establish the framework and show its effectiveness in conversation dynamic analysis.

As future work, we would like to explore other methods to tackle  limited data , such as incorporating prior knowledge and hierarchical parameter learning. Learning other parameters of the Hawkes process such as $\omega$, and, more generally, learning the decay kernel is an interesting direction for future research.

\bibliographystyle{abbrv}
\bibliography{bibfile}

\end{document}